\documentclass[11pt, draftcls, onecolumn]{IEEEtran}
\usepackage{amsmath,amssymb,bm,setspace,multirow,algorithm,algorithmic,epsfig,subcaption}

\newcommand{\scriptit}[1]{\scriptstyle{\mathsf{#1}}}
\newcommand{\argmax}{\operatornamewithlimits{argmax}}
\newcommand{\argmin}{\operatornamewithlimits{argmin}}

\title{Distributed Training of Deep Neural Network Acoustic Models for Automatic Speech Recognition}
\author{Xiaodong Cui, Wei Zhang, Ulrich Finkler, George Saon, Michael Picheny, David Kung\\
IBM Research AI\\ IBM T. J. Watson Research Center, Yorktown Heights, NY, 10598, USA \vspace{-1cm}}

\begin{document}

\maketitle

\begin{abstract}
The past decade has witnessed great progress in Automatic Speech Recognition (ASR) due to advances in deep learning. The improvements in performance can be attributed to both improved models and large-scale training data. Key to training such models is the employment of efficient distributed learning techniques. In this article, we provide an overview of distributed training techniques for deep neural network acoustic models for ASR. Starting with the fundamentals of data parallel stochastic gradient descent (SGD) and ASR acoustic modeling, we will investigate various distributed training strategies and their realizations in high performance computing (HPC) environments with an emphasis on striking the balance between communication and computation. Experiments are carried out on a popular public benchmark to study the convergence, speedup and recognition performance of the investigated strategies.\footnote{Accepted by IEEE Signal Processing Magazine. © 2020 IEEE.  Personal use of this material is permitted.  Permission from IEEE must be obtained for all other uses, in any current or future media, including reprinting/republishing this material for advertising or promotional purposes, creating new collective works, for resale or redistribution to servers or lists, or reuse of any copyrighted component of this work in other works.}
\end{abstract}

\vspace{-0.1cm}
\section{Introduction}
\label{sec:intro}

Deep Learning has had great impact across a broad variety of areas such as computer vision, natural language processing and automatic speech recognition (ASR) over the past several years. It is also the driving force behind the current Artificial Intelligence (AI) technologies that have achieved unprecedented success in a wide spectrum of applications.  ASR is one of the first areas that witnessed performance breakthroughs using deep learning techniques \cite{Hinton_DNNSPM}. Models that employ deep architectures are dominant in today's ASR systems that are ubiquitous in our everyday life, for instance, Google's voice search, Amazon's Alexa, Apple's Siri and IBM's Watson speech-to-text service. On some datasets, the word error rates (WERs) of high-performance ASR can even achieve human parity \cite{Saon_2017HumanParity}\cite{Xiong_ASRParity}.

The high performance of deep learning heavily relies upon large amounts of training data and high computational power. For instance, the amount of training speech data for ASR nowadays can easily reach thousands of hours, is often tens of thousands of hours, and some tasks even utilize as much as one million hours \cite{Parthasarathi_PetaAM}. Moreover, deep neural networks (DNNs) can have tens of millions of parameters or more. Without an appropriate distributed training strategy, training such models with deep architectures using this order of magnitude of data may take months to finish on a single GPU. Therefore, it is critical to investigate strategies that can speed up the training and reduce the turn-around time to an acceptable level. In this context, distributed learning has been demonstrated to be a very effective approach for training speedup. Over the years, various distributed machine learning strategies have been proposed and applied to different tasks. Distributed training of acoustic models as a branch of distributed machine learning has a lot in common with other domains such as computer vision in terms of algorithmic and system design. However, it also has characteristics unique to the ASR domain. Therefore, directly borrowing techniques from other domains may not give the best training performance for ASR. In the speech community, distributed acoustic modeling has been an active research topic. Notable work includes parameter server (PS) based synchronous training \cite{Seide_ParaSGD}\cite{Chen_BMUF}, PS based asynchronous training  \cite{Heigold_DistDNN}\cite{Heigold_DistSeq} and decentralized asynchronous training \cite{Zhang_ADPSGDSWB}\cite{Zhang_DDLASR}.

This tutorial will review commonly-used techniques for distributed training of acoustic models. We will first walk through some fundamentals of parallel stochastic gradient descent (SGD), high performance computing (HPC) architectures and deep acoustic modeling which are the foundation of the current large-scale distributed training. We will then present an in-depth investigation of several distributed strategies including synchronous/asynchronous and centralized/decentralized schemes and discuss their pros and cons. In particular, an emphasis is put on the interplay between computation and communication, which is the most important factor to design a successful distributed training strategy.  Experiments are carried out on the 2000-hour Switchboard (SWB2000) dataset \cite{Godfrey_SWB}\cite{Cieri_Fisher}, one of the most widely-used public benchmark datasets in the speech community \cite{Saon_2017HumanParity}\cite{Xiong_ASRParity}. We will conclude by analyzing the performance of various strategies. In particular, we will show that the asynchronous decentralized parallel SGD (AD-PSGD) algorithm recently proposed by IBM has achieved one of the best speedup performance to date on this dataset.

\vspace{-0.2cm}

\section{Optimization with SGD}
\label{sec:sgd}
\vspace{-0.1cm}
\subsection{Algorithm}
\label{sec:alg}

Most of the machine learning problems that employ DNNs are optimized using SGD. Suppose $\mathcal{X} \subseteq \mathbb{R}^{d_{x}}$ is the input space and $\mathcal{Y} \subseteq  \mathbb{R}^{d_{y}}$ is the output space of a supervised learning problem. We want to estimate a function $h$ with parameters $w$ that maps the input to the output
\vspace{-0.1cm}
\begin{align}
    h(w;x): \mathcal{X} \rightarrow \mathcal{Y}.  \vspace{-0.3cm}
\end{align}
A loss function $f(h(w;x),y)$ is used to measure the closeness between the prediction $h(w;x)$ and label $y$ where $x \in \mathcal{X}$ and $y \in \mathcal{Y}$. A risk function $F(w)$ given parameters $w$ is defined as the expected loss over the underlying joint distribution $p(x,y)$:
\vspace{-0.2cm}
\begin{align}
    F(w) = \mathbb{E}_{(x,y)}[f(h(w;x),y)] \triangleq \mathbb{E}_{\xi}[f(w,\xi)]   \label{eqn:risk}  \vspace{-0.3cm}
\end{align}
where $\xi \thicksim p(x,y)$ is a random variable on data $(x,y)$. We want to choose parameters $w$ that minimize $F(w)$
\vspace{-0.2cm}
\begin{align}
    w^{*} = \argmin_{w} F(w).  \label{eqn:minrisk}  \vspace{-0.3cm}
\end{align}

In practice, we only have access to a set of $n$ training samples $\{(x_{i},y_{i})\}_{i=1}^{n}$. Accordingly, we minimize the following empirical risk
\vspace{-0.2cm}
\begin{align}
     F(w) = \frac{1}{n} \sum_{i=1}^{n}f(w,(x_{i},y_{i}))         \label{eqn:emprisk} \vspace{-0.3cm}
\end{align}
where $\xi$ assumes the empirical data distribution.

When it comes down to large-scale optimization of DNNs in Eq.\ref{eqn:minrisk}, SGD is the dominant technique in deep learning due to its computational efficiency and competitive performance over other more complex optimization algorithms \cite{Bottou_SGD}. SGD solves the optimization problem of Eq.\ref{eqn:minrisk} iteratively. A basic SGD update formula is given by Eq.\ref{eqn:mbsgd}
\begin{align}
    w_{k+1} & = w_{k} - \alpha_{k}\cdot\left[\frac{1}{M}\sum_{m=1}^{M} \nabla f(w_{k};\xi_{k,m})\right]   \label{eqn:mbsgd}
\end{align}
where $w_{k}$ are the parameters after iteration $k$, $\alpha_{k}$ the learning rate and $\nabla f(w_{k};\xi_{k,m})$ the gradient evaluated at $w_{k}$ using the data samples denoted by the random variable $\xi_{k,m}$. There are $M$ samples randomly drawn from the whole $n$ training samples to form a so-called mini-batch. Their aggregated gradient is considered a ``noisy'' version of the true gradient $\nabla F(w)$ and hence a stochastic approximation \cite{Robbins_SA} of the deterministic gradient descent method. Therefore it is often referred to as the mini-batch based SGD and $M$ is the batch size. Besides the basic SGD algorithm given in Eq.\ref{eqn:mbsgd}, a variety of variants have been proposed in literature to improve its convergence properties where Adagrad \cite{Duchi_adagrad}, Adam \cite{Kingma_adam} and Nesterov acceleration \cite{Nesterov_NAG} are among the most notable ones. These SGD variants have found varied degrees of success in different applications. In this paper, for the tutorial purpose and for the ease of discussion, we will focus on the basic SGD form in Eq.\ref{eqn:mbsgd}.

\vspace{-0.3cm}

\subsection{Training Strategies}
\label{sec:strageties}

Strategies of distributed machine learning can be broadly categorized into the follow families:

\paragraph{data parallel vs. model parallel}
Data parallelism distributes mini-batches of data onto a number of learners (e.g. a GPU or CPU) with each learner having a copy of the model. The computation is carried out on each learner in parallel using the data before the system aggregates the statistics in some fashion. Sometimes, the model is too large to fit into the memory of one single learner. Under this condition, model parallelism is used to split the model across the learners with each learner only having a partial model. The output of one learner is used as the input of another learner to conduct the computation of the full model. Although model parallelism is used in some scenarios \cite{Seide_ParaSGD}, data parallelism is the dominant distributed strategy in practice in today's distributed learning community \cite{Jia_ImageNet4mins}\cite{Chen_BMUF}\cite{Heigold_DistDNN}.

\paragraph{single node vs. multiple nodes}

A node refers to a physical device (e.g. a server) in a computer network. Distributed training on a single node is a straightforward setup where all learners (e.g. GPUs) stay within one machine with a shared memory. The communication among the learners is reliable and it can give a moderate speedup. Therefore, it is still a reasonable option for distributed training of DNNs. Obviously it has the limitation of the capability of scaling out and the number of learners is limited to the hardware configuration of the machines. Multiple-node distributed training has a number of machines (nodes) to form a cloud or cluster where learners use shared memory for local communication within the node and message passing over the network for internode communication. It has become the norm for recent large-scale distributed training.

\paragraph{centralized vs. decentralized}

Centralized distributed training relies on a central PS and all learners only communicate with the PS for model updates. Decentralized setting has no PS and all learners form a network of certain topology (e.g. a ring). All learners are in an equal position in terms of computation and communication. Centralized distributed training is commonly used in practice in many of the machine learning applications. However, it imposes a large communication bottleneck to the PS as the hub when the number of learners is large. Decentralized distributed training, on the other hand, has an advantage on the communication bandwidth and is becoming more and more popular.

\paragraph{synchronous vs. asynchronous}

Distributed training algorithms can be run in synchronous or asynchronous mode. Under synchronous mode, gradient computation and model update are synchronized with each other. Model update is carried out after all learners finish local gradient computation. Under asynchronous mode, however, the synchronization between the gradient computation and model update is removed and learners receive their mini-batches as needed depending on their computation and communication speed, which significantly reduces the idle time relative to the synchronous mode.

In what follows, we will focus on data-parallel multiple-node distributed acoustic modeling since it is the most popular distributed setting nowadays. We will present an in-depth investigation of its behavior in centralized/decentralized configurations under synchronous/asynchronous modes.

\vspace{-0.3cm}

\subsection{HPC Architecture}
\label{sec:hpc}

Distributed computing aims at parallelizing execution among independent computational resources (e.g. processors). A parallel program consists of three components: (1) The parallel part that can be carried out in parallel, for example, gradient computation. (2) The sequential part that can not be parallelized. For example, the summation of gradients or models can only be accomplished after the gradients or models are in place. (3) The communication portion that passes information between computational resources, for instance, gradient/weight transfer among learners. Assuming communication cost is zero, Amdahl's law states that the speedup of a parallel program is bounded by $\frac{1}{1-p}$, where $p$ is the portion of the parallelizable part of the program. For the SGD algorithm, the parallel part dominates the sequential part. Hence $p$ is very close to 1 and very high speedups can be achieved in principle. As a result, the fundamental limiting factor to achieve linear speedup in this case is the communication cost.

\begin{figure}[tbh]
   \centering
   \includegraphics[width=6cm, height=5cm]{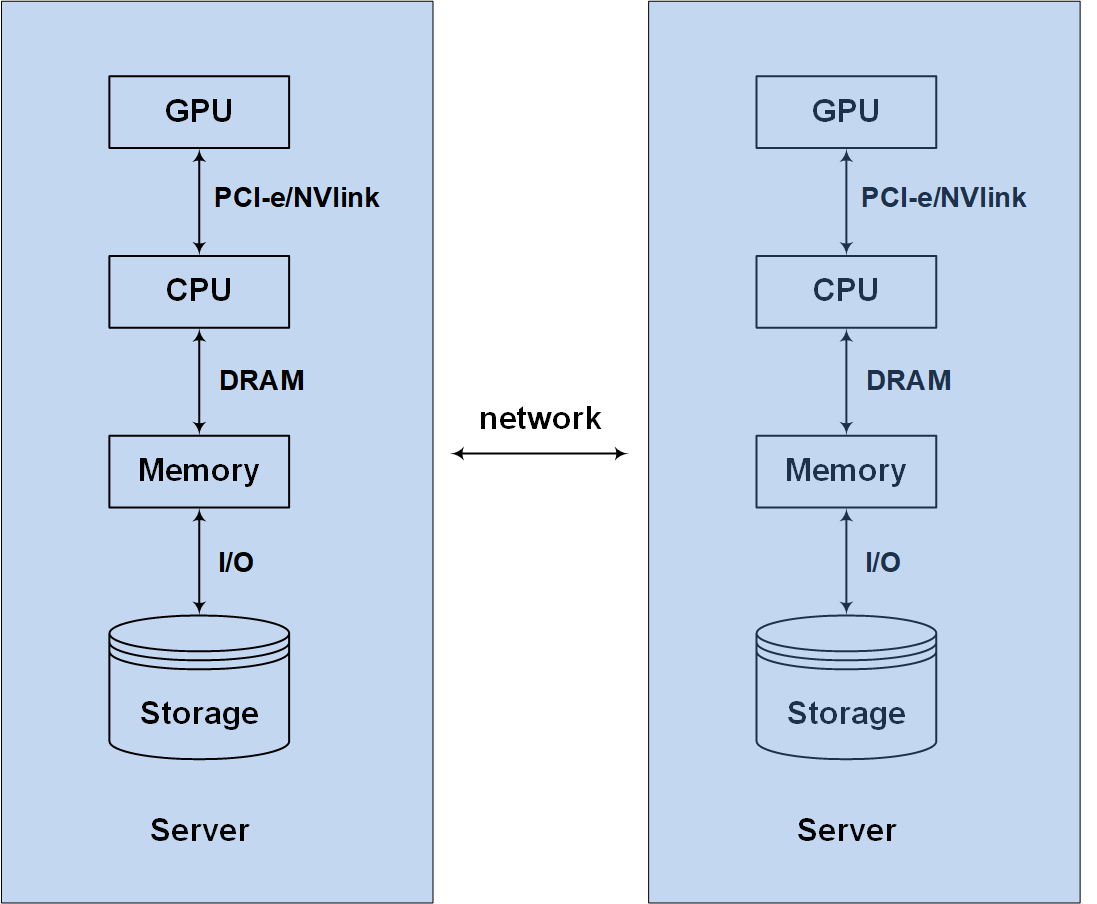}
   \caption{Components and data flow in an HPC environment for distributed training.}\label{fig:hpc}\vspace{-0.7cm}
\end{figure}

A typical distributed training system, as depicted in Fig.\ref{fig:hpc}, has the following hardware components: data storage, memory,  processing units (CPU/GPU) and the network. Data communication follows this path: Each learner loads input data from data storage under storage bandwidth constraint to main memory. It conducts data-preprocessing using the CPU before sending it to GPU via CPU-GPU databus for model training. When the gradient computation is finished, the gradients are sent to the PS or other learners under the constraint of network bandwidth. Typical bandwidth of storage systems ranges from 1-10 MB/s (network file system) to 100MBs/s (hard disk drives, HDD), to 300-500MB/s (solid state drive, SSD), to several GB/s (non-volatile memory express (NVMe) SSDs). Typical main memory bandwidth is of several tens of GB/s. Typical CPU-GPU databus bandwidth ranges from 16GB/s one way (e.g. PCI-e 3rd gen) to 50GB/s one way (e.g. IBM Power 9 Nvlink). Typical network bandwidth ranges from 100MB/s (e.g. 1Gb Ethernet) to 10GB/s (e.g. 100Gb Ethernet, RDMA). A high-end HPC cluster (e.g. SUMMIT supercomputer) usually is equipped with NVMe SSDs, NVlinks and 100Gb Ethernet (or higher) to enable fast communication. On the computation side, the key deciding factor is FLOPS. Typical high-end CPUs run at hundreds of GFLOPS to 1 TFLOPS and typical high-end GPUs run at 10 TFLOPs or higher. System programmers use concurrent programming, usually achieved by multi-threading, to overlap communication with computation. In synchronous mode, data-loading and data-processing can be overlapped with gradient computation. In asynchronous mode, all communication paths (i.e. data-loading, data preprocessing, CPU-GPU data transfer, and gradients/weights transfer) can be overlapped with gradient computation. Ideally, we wish to pursue perfect overlap between communication and computation. In practice, the communication that cannot be overlapped by computation is the limiting factor in achieving linear speedup. One of the important tasks in designing distributed learning is to maximize the overlap between the two.   \vspace{-0.5cm}

\section{Acoustic Modeling in ASR}
\label{sec:am}

\begin{figure}[tbh]
   \centering
   \includegraphics[width=9.5cm, height=2.2cm]{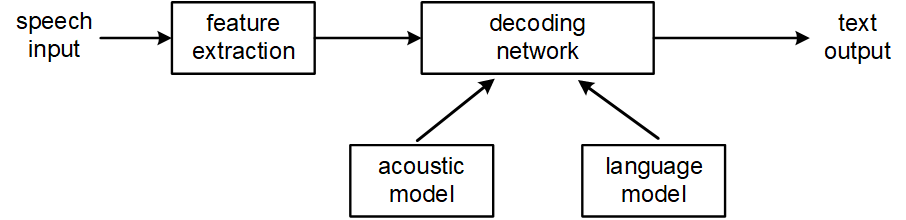}
   \caption{Components of an ASR system.}\label{fig:asr}\vspace{-0.7cm}
\end{figure}

Suppose $X = \{x_{1}, x_{2}, \cdots, x_{m}\}$ is a sequence of acoustic features and $W = \{w_{1}, w_{2}, \cdots, w_{n}\}$ is a sequence of words. ASR systems find the most likely word sequence $W$ given the observed acoustic feature sequence $X$:
\begin{align}
   W^{*} = \argmax_{W} P(W|X) = \argmax_{W} \frac{P(X|W)P(W)}{P(X)}  \label{eqn:asr}
\end{align}
This is illustrated in Fig.\ref{fig:asr}. There are four major components in an ASR system.  A feature extractor converts a speech waveform into a sequence of acoustic feature vectors (e.g. logMel features). An acoustic model computes the probability that a particular word sequence can produce the observed acoustic features, $P(X|W) = P(x_{1}, x_{2}, \cdots, x_{m} | w_{1}, w_{2}, \cdots, w_{n})$.  A language model computes the probability of a particular word sequence, $P(W)= P(w_{1}, w_{2}, \cdots, w_{n})$. Finally, there is a decoder which searches for the best word sequence $W^{*}$ that maximizes Eq.\ref{eqn:asr}.

In this article, we consider DNN acoustic models based on a hidden Markov model (HMM) structure, often referred to as DNN-HMM. HMMs are probabilistic automata that are commonly used for modeling sequences of variable length such as speech. Under this structure, words are represented as strings of speech sounds, called “phones”. Each phone is represented by an HMM with numerous states. Since a phone can be affected by its context (i.e. phones immediately before and after it), modern ASR systems use the so-called context-dependent (CD) phones. The HMM states of each CD phone are clustered using a decision tree based on their acoustic similarity. This gives rise to a large number of CD HMM states as fine-grained phone classes for  acoustic modeling. In DNN-HMMs, the output of the DNNs after the topmost softmax represents a set of posterior probabilities corresponding to each of the CD HMM states \cite{Dahl_CDDNN}. Modern acoustic models use either CNNs or recurrent networks the most popular of which are LSTMs \cite{Saon_2017HumanParity}\cite{Xiong_ASRParity}. Training of acoustic models involves optimizing the DNNs under an appropriate objective function. In ASR, frame-based cross-entropy (CE) \cite{Hinton_DNNSPM} and sequence-based loss functions such as state-level minimum Bayes error (sMBR) \cite{Kingsbury_Seqtrain} are commonly used for optimization. In recent years, end-to-end (E2E) ASR systems based on connectionist temporal classification (CTC) \cite{Graves_CTC} or encoder-decoder \cite{Chan_LAS} structures are also drawing attention in the speech community. This tutorial will use the LSTM-based DNN-HMM acoustic models trained with the CE criterion to investigate the distributed training strategies in ASR as they are most representative. But the techniques presented are applicable in principle to other models and training criteria as well.

\section{Distributed Training of Acoustic Models}
\label{sec:dist_am}

\subsection{Unique Characteristics}
\label{sec:char}

DNN acoustic models have distinct characteristics from other domains (e.g. computer vision) in terms of distributed training. A DNN-HMM acoustic model typically has a softmax layer with a large number of output CD phone classes \cite{Xiong_ASRParity}\cite{Saon_2017HumanParity}\cite{Dahl_CDDNN} which are usually on the order of magnitude of 10,000. Moreover, the distribution of speech samples across phone classes is hugely uneven. In addition, the input feature space is relatively more structured for speech signals. Therefore, the DNN acoustic models are usually shallower than those in vision. Table \ref{tab:modelcomp} shows the configurations of two representative DNN models for speech and vision tasks, a 50-layer ResNet model for the ImageNet image recognition task and a 6-layer LSTM model for the Switchboard ASR task. The ResNet model has a smaller model size even with more convolutional layers due to parameter sharing and local connectivity. However, the convolutional operation is computationally expensive, which results in a longer computing time for each batch. On the other hand, the LSTM model for the speech task has a bigger model but the computation is faster. Overall, the optimization of an acoustic model has a lighter computational load but a heavier communication load relative to the vision models. This \textbf{high communication/computation ratio} imposes a major challenge for distributed training of DNN acoustic models.

\begin{table}[tbh]
\centering
\begin{tabular}{l|c|c|c|c|c} \hline
                  &     model      &      layers        &    output     &     size      &   computation/batch     \\ \hline\hline
      vision      &    ResNet      &        50          &     1,000     &  $\sim$ 100MB   &  $\sim$ 0.18 sec    \\ \hline
      speech      &    LSTM        &         8          &    32,000     &  $\sim$ 165MB   &  $\sim$ 0.07 sec    \\ \hline
\end{tabular}\vspace{0.2cm}
\caption{Model comparison between speech and computer vision. A ResNet model is used for the ImageNet task and an LSTM model is used for the SWB2000 task. The last column shows the seconds used to compute a batch of size 32 on a P100 GPU.}\label{tab:modelcomp}\vspace{-0.7cm}
\end{table}

\subsection{Centralized Distributed Training}
\label{sec:centralized}

\begin{figure}[ht]
  \centering
  \begin{subfigure}{.4\textwidth}
      \centering
      \includegraphics[width=5cm, height=4.5cm]{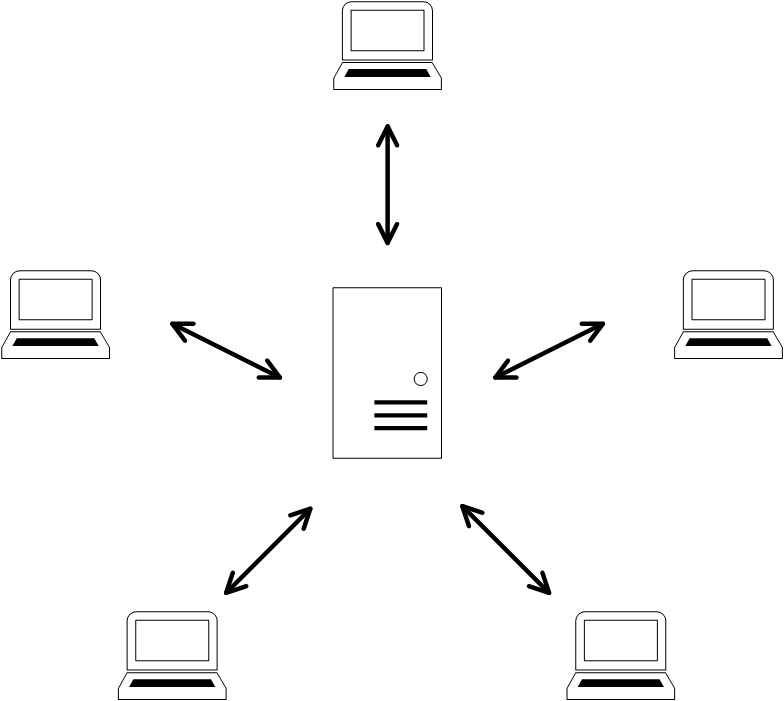}
      \caption{centralized}\label{fig:csgd}
  \end{subfigure}
  \hspace{1cm}
  \begin{subfigure}{.4\textwidth}
      \centering
      \includegraphics[width=5cm, height=4.5cm]{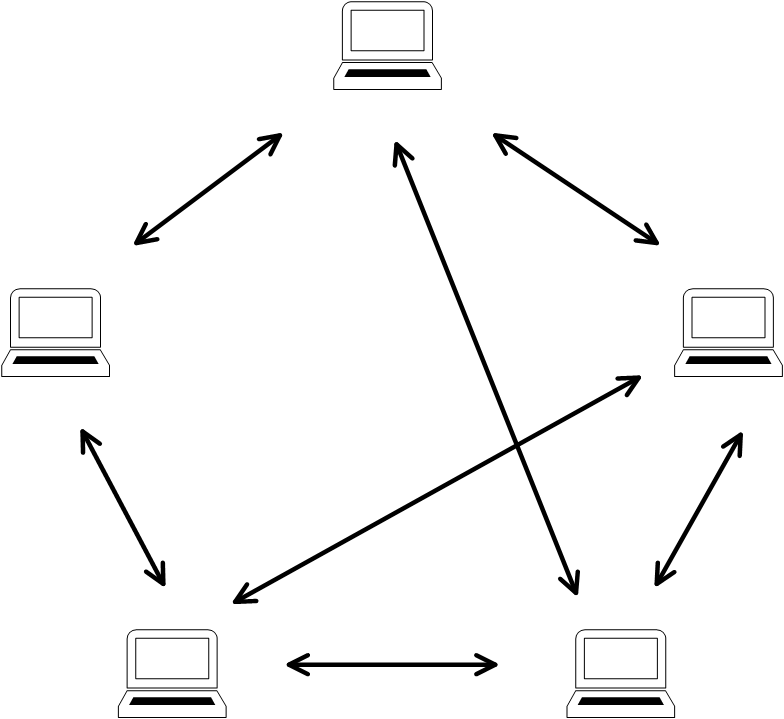}
      \caption{decentralized}\label{fig:dsgd}
  \end{subfigure}
  \caption{Centralized distributed training with a parameter server (left) and decentralized distributed training without a parameter server where communication takes place among the learners (right).}\vspace{-0.7cm}
\end{figure}

Centralized distributed training has a PS serving as a hub in the network, as illustrated in Fig.\ref{fig:csgd}. The PS has the global view of the model and all learners only communicate with the PS. Each learner pulls the model from the PS. The computation of gradients is carried out local on each learner, after which the gradients are pushed back to the PS. The PS collects the statistics from the learners and updates the model accordingly, depending on whether the implementation is synchronous or asynchronous.

Consider the SGD iteration given in Eq.\ref{eqn:mbsgd}. Suppose each mini-batch $M$ is evenly split onto $L$ learners with each learner $l$ having a batch size of $M_{l}=M/L$.

\subsubsection{Synchronous Parallel SGD}

Under synchronous parallel SGD, each learner $l$ pulls model $w_{k}$ from the PS, computes the local gradient \vspace{-0.2cm}
\begin{align}
   g^{(l)}(w_{k}, \xi_{k}) = \frac{1}{M_{l}}\sum_{m=1}^{M_{l}} \nabla f(w_{k};\xi_{k,m})   \label{eqn:ccsgd_g} \vspace{-0.2cm}
\end{align}
and then pushes $g^{(l)}(w_{k}, \xi_{k})$ back to the PS.

The PS will wait until all $L$ learners finish their local gradient computation to aggregate them up for model update \vspace{-0.2cm}
\begin{align}
    w_{k+1} = w_{k} - \alpha_{k} \cdot \left[\frac{1}{L}\sum_{l=1}^{L} g^{(l)}(w_{k}, \xi_{k})\right] \label{eqn:ccsgd_w} \vspace{-0.2cm}
\end{align}
Note that all learners use the same copy of the model $w_{k}$ for the computation of $g^{(l)}(w_{k}, \xi_{k})$ which is consistent with the global copy of the model residing on the PS. Eqs. \ref{eqn:ccsgd_g} and \ref{eqn:ccsgd_w} give the same update as Eq.\ref{eqn:mbsgd}. Synchronous parallel SGD is the most reliable implementation of SGD in a distributed setting and shares the same convergence property. Synchronous SGD may suffer from the well-known ``straggler" problem because the PS has to wait for the slowest learner. This synchronization cost limits the overall speedup. Nevertheless, because of its simplicity, centralized synchronous distributed training of acoustic models is still popular in the speech community. Representative work includes \cite{Seide_ParaSGD}\cite{Chen_BMUF} form Microsoft, \cite{Parthasarathi_PetaAM} from Amazon, \cite{Hannun_Deepspeech} from Baidu.  The global model updates can be conducted either through gradient aggregation \cite{Seide_ParaSGD}\cite{Hannun_Deepspeech} or model averaging \cite{Chen_BMUF}.  One of the notable work among them is a strategy based on blockwise model-update filtering (BMUF) \cite{Chen_BMUF} proposed by Microsoft, which is a variant of synchronous SGD. Under BMUF, data is partitioned into blocks. Each worker updates its local model in parallel using SGD. Instead of performing direct model averaging, the global model is updated using block-level stochastic optimization by synchronizing local models from all learners based on the block momentum. Good performance has been reported by Microsoft on large-scale distributed acoustic modeling and has also been used by Amazon to train DNN acoustic models using one million hours of data \cite{Parthasarathi_PetaAM}.

\subsubsection{Asynchronous Parallel SGD}

Under asynchronous parallel SGD, each learner $l$ pulls model $\hat{w}_{k}$ from the PS and computes the local gradient \vspace{-0.2cm}
\begin{align}
   g^{(l)}(\hat{w}_{k}, \xi_{k}) = \frac{1}{M_{l}}\sum_{m=1}^{M_{l}} \nabla f(\hat{w}_{k};\xi_{k,m})   \label{eqn:casgd_g} \vspace{-0.2cm}
\end{align}
then pushes $g^{(l)}(\hat{w}_{k}, \xi_{k})$ back to the PS.

The PS will update the model right after receiving the gradient from learner $l$  \vspace{-0.2cm}
\begin{align}
    w_{k+1} = w_{k} - \alpha_{k} \cdot g^{(l)}(\hat{w}_{k}, \xi_{k}) \label{eqn:casgd_w}  \vspace{-0.2cm}
\end{align}
In this case there may exist inconsistency between the model $w_{k}$ on the PS and the model $\hat{w}_{k}$ pulled by learner $l$ for the computation of its local gradient  \vspace{-0.2cm}
\begin{align}
\hat{w}_{k} = w_{k-\tau}, \ \ \ \ \tau \geq 0  \vspace{-0.2cm}
\end{align}
This is because the model on the PS may have been updated by other learners while learner $l$ is still computing its local gradients. This inconsistency is often referred to as staleness. Asynchronous parallel SGD, with the synchronization removed between the learner and server, can significantly reduce the idle time and improve the speedup. It does not have the ``straggler" problem and can automatically balance the workload among the fast and slow learners. Nevertheless, the incurred staleness may hurt convergence and eventually the learning performance.

In the speech community, centralized asynchronous distributed training of acoustic models has also been being used \cite{Heigold_DistDNN}\cite{Heigold_DistSeq}. Downpour SGD based on the DistBelief framework \cite{Dean_LargeDistNN}, proposed by Google, is a representative application. Distributed training ranging from multilingual acoustic modeling \cite{Heigold_DistDNN} to sequence acoustic modeling \cite{Heigold_DistSeq} has been reported using this strategy to deliver good performance. Downpour SGD is a variant of asynchronous SGD. It divides the data into blocks and distributes them onto multiple learners. Each learner keeps a local copy of the model and uses it to carry out computation in parallel. They independently push the updates to the PS, which keeps the current state of model, and then pull the updated model from PS. The PS itself is sharded across multiple machines and each shard is only responsible for updating part of the model. Downpour SGD introduces asynchrony to both local learners and PS shards and has been shown to be robust to machine failure during training. But centralized asynchronous distributed training may be bothered by the potential large staleness issue in general and it is challenging to have a good parallelization efficiency and convergence behavior.


\subsection{Decentralized Distributed Training}
\label{sec:decentralized}

Centralized distributed training relies on a PS to communicate with all the learners. All the communication takes place between the server and the learners and there is no communication among the learners. This introduces a high communication cost at the PS, which is proportional to the number of learners, and will eventually hurt the scaling of the training. A PS does not necessarily need to be realized as a physical server or sharded servers \cite{Dean_LargeDistNN}. It can also be conceptually realized via the allreduce operation \cite{Patarasuk_Allreduce} based on message passing. In HPC terminology, an operation is a ``reduce" operation if it is commutative and associative (e.g. summation). Allreduce is the operation that reduces all the elements and broadcasts the reduction results to each participant. Decentralized distributed training does not require a centralized server to support the communication. Fig.\ref{fig:dsgd} shows how decentralized SGD works. The learners form a network by connecting with each other following some topology (e.g. a ring). Each learner keeps a local copy of the model and carries out gradient computation and model updates locally. The updated local model is then propagated to other learners in the network typically via model averaging.

Consider the centralized data parallel SGD setting of Eq.\ref{eqn:mbsgd} under synchronous mode. Suppose each learner pulls the model from the PS, evaluates the gradient using batch size $M_{l}$ and updates the model locally \vspace{-0.2cm}
\begin{align}
    w^{(l)}_{k+1} & = w_{k} - \alpha_{k}\cdot\left[\frac{1}{M_{l}}\sum_{m=1}^{M_{l}} \nabla f(w_{k};\xi_{k,m})\right] \vspace{-0.2cm}
\end{align}
Then we average models across all the learners \vspace{-0.2cm}
\begin{align}
    w_{k+1} = \frac{1}{L}\sum_{l=1}^{L}w^{(l)}_{k+1}  =  w_{k} - \alpha_{k}\cdot\left[\frac{1}{M}\sum_{m=1}^{M} \nabla f(w_{k};\xi_{k,m})\right]  \label{eqn:c2de} \vspace{-0.2cm}
\end{align}
which shows that given the basic SGD in Eq.\ref{eqn:mbsgd} one-step model averaging and gradient averaging are equivalent. Eq.\ref{eqn:c2de} also provides a way to create a decentralized implementation of centralized SGD where all learners form a ring and the PS is replaced by a global model averaging carried out by the allreduce operation using reduction (sum) followed by broadcast.

In general, the mathematical model of data parallel decentralized SGD is given by Eq.\ref{eqn:dsgd}: \vspace{-0.2cm}
\begin{align}
    \mathbf{W}_{k+1} = \mathbf{W}_{k} \cdot \mathbf{T} - \alpha_{k} \cdot g(\mathbf{\Phi}_{k},\bm{\xi}_{k})  \label{eqn:dsgd} \vspace{-0.2cm}
\end{align}
where, for $l = 1, \dots, L$, \vspace{-0.2cm}
\begin{itemize}
   \item $\mathbf{W}_{k} = [w^{\scriptit{(1)}}_{k}, \dots, w^{\scriptit{(l)}}_{k}, \dots, w^{\scriptit{(L)}}_{k}]$ is a matrix with each column containing model parameters in each learner $l$ at iteration $k$
   \item $\mathbf{T}$ is a mixing matrix for the model averaging pattern among learners given a network topology
   \item $\mathbf{\Phi}_{k} = [\hat{w}^{\scriptit{(1)}}_{k}, \dots, \hat{w}^{\scriptit{(l)}}_{k}, \dots, \hat{w}^{\scriptit{(L)}}_{k}]$ is a matrix with each column containing model parameters used for computing gradient in each learner $l$ at iteration $k$
   \item $\bm{\xi}_{k} = [\xi^{\scriptit{(1)}}_{k}, \dots, \xi^{\scriptit{(l)}}_{k}, \dots, \xi^{\scriptit{(L)}}_{k}]$ is a matrix with each column containing indexing random variables for mini-batch samples used for computing gradients in each learner $l$ at iteration $k$
   \item $g(\mathbf{\Phi}_{k},\bm{\xi}_{k}) = [\frac{1}{M_{1}}\sum_{m=1}^{M_{1}} \nabla f(\hat{w}^{\scriptit{(1)}}_{k};\xi^{\scriptit{(1)}}_{k,m}), \dots, \frac{1}{M_{L}}\sum_{m=1}^{M_{L}} \nabla f(\hat{w}^{\scriptit{(L)}}_{k};\xi^{\scriptit{(L)}}_{k,m})]$ is a matrix with each column containing gradients computed in each learner $l$ at iteration $k$
\end{itemize}
The first term on the right in Eq.\ref{eqn:dsgd} describes the communication pattern among learners while the second term depends on gradient computation on each learner. The two terms can be evaluated concurrently. Each learner keeps a local model and computes the gradients locally. Meanwhile, the local model is also averaged with other learners in the network through the mixing matrix. If the computation takes a longer time than the communication, the first term can be entirely overlapped by the second term. Eq.\ref{eqn:dsgd} can also be carried out in synchronous \cite{Lian_DecentSGD}\cite{Zhang_ADPSGDSWB} or asynchronous \cite{Lian_ADPSGD}\cite{Zhang_ADPSGDSWB} mode. In synchronous mode, the model update has to hold until the two terms are both in place. In asynchronous mode, the model update takes place whenever a learner finishes its local computation. By introducing various synchronization mechanisms between the two terms, Eq.\ref{eqn:dsgd} can cover a broad variety of decentralized training strategies.

It is also worth pointing out that although it is suitable for theoretical analysis of convergence behaviors of a training strategy, Eq.\ref{eqn:dsgd} does not reflect its communication cost which mainly includes the time of data transfer from storage and memory, model/gradient transfer between CPU and GPU and model/gradient averaging among learners.

Model averaging in the decentralized SGD is indicated by the mixing matrix $\mathbf{T}$ which is typically chosen as \textbf{doubly stochastic matrices}. A matrix $\mathbf{T}=(t_{ij})$ is called a doubly stochastic matrix if $t_{ij} \in [0,1]$ and $\sum_{i}t_{ij}=\sum_{j}t_{ij}=1$. For instance,
\[
\mathbf{T}_{1} = \begin{bmatrix}
    \frac{1}{3} & \frac{1}{3} &    0        &    0        &     0     &     0         &   \frac{1}{3}  \\
    \frac{1}{3} & \frac{1}{3} & \frac{1}{3} &    0        &     0     &     0         &       0        \\
    0           & \frac{1}{3} & \frac{1}{3} & \frac{1}{3} &     0     &     0         &       0        \\
      \cdots    &   \cdots    &  \cdots     &  \cdots     &   \cdots  & \cdots        &   \cdots  \\
    \frac{1}{3} &    0        &    0        &    0        &     0     &  \frac{1}{3}  &  \frac{1}{3}
\end{bmatrix}  \ \ \ \
\mathbf{T}_{u} = \begin{bmatrix}
    \frac{1}{L} & \frac{1}{L} & \cdots &  \cdots &  \cdots &  \frac{1}{L} & \frac{1}{L} \\
    \frac{1}{L} & \frac{1}{L} & \cdots &  \cdots &  \cdots &  \frac{1}{L} & \frac{1}{L} \\
    \frac{1}{L} & \frac{1}{L} & \cdots &  \cdots &  \cdots &  \frac{1}{L} & \frac{1}{L} \\
     \cdots    &   \cdots    & \cdots &  \cdots &  \cdots &   \cdots    &  \cdots     \\
    \frac{1}{L} & \frac{1}{L} & \cdots &  \cdots &  \cdots &  \frac{1}{L} & \frac{1}{L}
\end{bmatrix}
\]
where $\mathbf{T}_{1}$ represents a model averaging scheme in which each learner averages its local models with its immediate left and right neighbors in a ring. $\mathbf{T}_{u}$ represents the scheme where local models of all learners are averaged. Treating $\mathbf{T}$ as a transition matrix of a Markov chain, if its represented chain is irreducible and aperiodic, it has a stationary uniform distribution $\mathbf{T}_{u}$: $\mathbf{T}^{n} \rightarrow \mathbf{T}_{u}$ when $n \rightarrow \infty$. It indicates that with sufficient rounds of model averaging under $\mathbf{T}$, local models of all learners will reach consensus, which is the average of all local models.

In recent years, decentralized parallel SGD has been theoretically shown to be equally good in terms of convergence rate as the conventional SGD \cite{Lian_DecentSGD}\cite{Lian_ADPSGD}. Communication-wise, decentralized parallel SGD is advantageous over the centralized strategies as it removes the communication barrier on the PS. Recent work from IBM \cite{Zhang_ADPSGDSWB}\cite{Zhang_DDLASR} has demonstrated good performance in both speedup and WER using asynchronous decentralized strategies in large-scale acoustic modeling. In particular, a hybrid distributed setting was proposed in \cite{Zhang_DDLASR} that combines synchronous and asynchronous modes under the same decentralized parallel SGD framework. Asynchronous decentralized strategies are also found to tolerate larger batch sizes relative to the centralized strategies.

\subsection{Improving Training Efficiency}
\label{sec:eff}

When designing a distributed training strategy based on SGD, batch size and communication bandwidth are two critical factors to consider in practice for training efficiency.

For data parallel SGD, the speedup is roughly proportional to the batch size under the constraints of GPU memory and model size. The more data we can parallelize in each batch the faster the training is. It is difficult to increase the number of learners while maintaining a high percentage GPU usage if the batch size is not sufficiently large. However, it is often observed that too large a batch size may hurt the convergence of SGD and eventually the performance of the model \cite{Zhang_DDLASR}. Therefore, effectively increasing the batch size without compromising performance has been actively investigated over the years in the distributed training community \cite{Jia_ImageNet4mins}\cite{Zhang_ADPSGDSWB}. To deal with the batch size issue, learning rate warm-up is often used.  A common practice is that a large batch is learned with a large learning rate, which are roughly in proportion, but the large learning rate is achieved by gradually scaling up from a small learning rate. This strategy usually gives good performance in practice. It is also observed that larger batch sizes are possible in decentralized asynchronous SGD with partial model averaging \cite{Zhang_DDLASR}.

Bandwidth indicates how much data can be communicated per second.  In parallel SGD, we always hope for a perfect overlap between communication and computation to minimize the training time. The communication takes place when data is copied from storage to memory, models/gradients are transferred between CPU and GPU and models/gradients are averaged among learners. The computation mainly involves gradient evaluation. We want to push computation-heavy operations to GPUs while reducing the communication cost. DNN models with large number of parameters require high communication bandwidth and may become the eventual bottleneck of the whole distributed training. Specialized to distributed acoustic modeling, the issue may become even more severe due to its high communication/computation ratio. First of all, loading features and labels from storage to memory may affect the training efficiency given the low bandwidth between the two. A typical way to deal with it is to run data loaders in multiple processes in parallel to pipeline data loading and perform online feature expansion if necessary. In terms of model/graident transfer, a broad variety of communication-reduction techniques have been proposed. Gradient compression approaches such as gradient quantization \cite{Seide_1bit}\cite{Alistarh_QSGD} and gradient sparsification \cite{Aji_SparseSGD} are used to reduce the required communication bandwidth. Partial model averaging instead of global model averaging is another way to reduce the communication cost \cite{Lian_DecentSGD}\cite{Zhang_ADPSGDSWB}.

When training acoustic models under discriminative sequence criteria, additional care needs to be taken with issues on storage, communication and computation in terms of training efficiency. The hypothesis space in the discriminative objective function is represented by lattices \cite{Kingsbury_Seqtrain} which take significant amount of storage space. As a result, the data loading is more time-consuming compared to the CE training. On the computation side, the gradient evaluation involves the forward and backward algorithm running on lattices, which typically takes place on CPUs as it is nontrivial to express it in an efficient form of matrix multiplication suitable for GPUs. For large-scale distributed training, shallow lattices of low density are usually used to reduce the required storage space and speed up the communication and computation \cite{Parthasarathi_PetaAM}.

\section{Experiments}
\label{sec:exp}

In this section, we evaluate various distributed deep acoustic model training strategies on SWB2000.  The first set of experiments are designed to compare their performance in convergence, speedup and WER for the pedagogical purpose.

\textbf{Dataset}  \ \  SWB2000 is a well-established public benchmark for ASR evaluation \cite{Hinton_DNNSPM}\cite{Saon_2017HumanParity}\cite{Xiong_ASRParity}. The dataset consists of 1,975 hours of audio data among which 10 hours of audio is used for the heldout set for training.  WERs are evaluated on the Hub5 2000 evaluation set which is composed of two parts: One is the 2.1-hour switchboard (SWB) data and the other is the 1.6-hour callhome (CH) data.

\textbf{Model}  \ \  The acoustic model is a DNN-HMM with a bi-directional LSTM architecture. There are 6 LSTM layers and each LSTM layer contains 1,024 cells with 512 in each direction. On top of the LSTM layers, there is a linear bottleneck layer with 256 hidden units followed by a softmax output layer with 32,000 units corresponding to the CD-HMM states. The LSTM is unrolled 21 frames and trained with non-overlapping feature subsequences of that length. The input is a 260-dimensional vector consisting of a speaker-adapted acoustic feature based on perceptual linear prediction (PLP) \cite{Hermansky_PLP} (40-dim), a speaker embedding vector \cite{Dehak_ivec} (100-dim) and a logMel feature with its delta and double-delta \cite{Saon_2017HumanParity} ($3\!\times\!40$-dim). The language model is trained using publicly available text data from a wide variety of sources. The final LM used for decoding has 36M 4-grams with a vocabulary of 85,000 words.

\textbf{Baseline} \ \  We establish the baseline by training the acoustic model using SGD on a single P100 GPU without parallelization. The batch size is 256. Following the input dimensionality and LSTM unroll length, a batch is a tensor of size $260\!\times\!21\!\times\!256$. The initial learning rate is 0.1 which is annealed by $\frac{1}{\sqrt{2}}$ every epoch after 10th epoch. The training finishes after 16 epochs. The WERs are 7.5\% on SWB and 13.0\% on CH, one of the best results under CE training on this dataset. This is a well-tuned training recipe where the learning rate scheduling and batch size are optimized towards the best WERs.

\textbf{HPC Setting} \ \ We use a cluster of 4 x86 servers with 2 7-core Intel Xeon E5-2680 2.40GHz CPUs and 1TB main memory per server. Each server has 4 P100 GPUs. Therefore, there are 16 GPUs in total. Servers are connected with 100 Gbit/s Ethernet. On each server, CPU and GPU communicate via PCI-e Gen3 bus with a 16GB/s peak bandwidth in each direction. The audio data is first converted to input features and labels in the HDF5 format and stored locally on NVMe on each server. The connection among learners has a ring topology. In case of allreduce, it is implemented using NCCL \cite{NCCL}.  Each learner has 2 concurrent processes on CPUs to load the input features and labels. The data loader generates 21-frame subsequences for unrolled LSTM and expands the delta and double delta of logMel features on the fly, which overlaps with the gradient evaluation on GPUs. No gradient compression is used in communication.

\textbf{Training Strategies} \ \ We implement and compare three distributed training strategies. (1) \textbf{SC-PSGD}: synchronous decentralized parallel SGD with allreduce for model averaging (mixing matrix $\mathbf{T}_{u}$). This is equivalent to a synchronous centralized parallel SGD where the PS  operations are replaced with a reduction-then-broadcast allreduce operation. (2) \textbf{SD-PSGD}: synchronous decentralized parallel SGD in which each learner averages its model with its left and right neighbors (mixing matrix $\mathbf{T}_{1}$).  This model averaging pattern can reduce the communication cost compared to allreduce. (3) \textbf{AD-PSGD}: asynchronous decentralized parallel SGD where local gradient computation and model update running concurrently with model averaging with its left and right neighbors (mixing matrix $\mathbf{T}_{1}$).  Asynchronous centralized parallel SGD is not included in the experiments as it is known to be hard to train and gradually loses popularity to other strategies.

\begin{figure}[ht]
  \centering
  \begin{subfigure}{.4\textwidth}
      \centering
      \includegraphics[width=\textwidth, height=5cm]{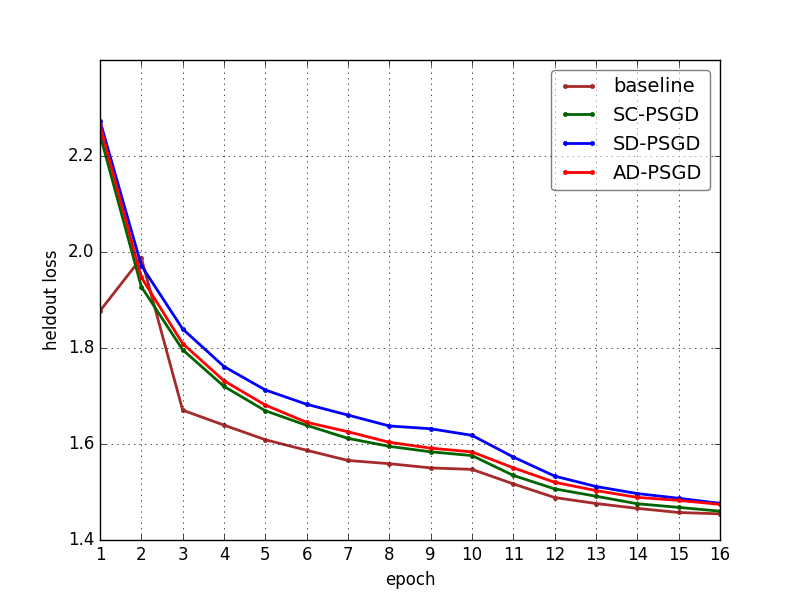}
      \caption{heldout loss}
  \end{subfigure}
  \hspace{0.5cm}
  \begin{subfigure}{.4\textwidth}
      \centering
      \includegraphics[width=\textwidth, height=5cm]{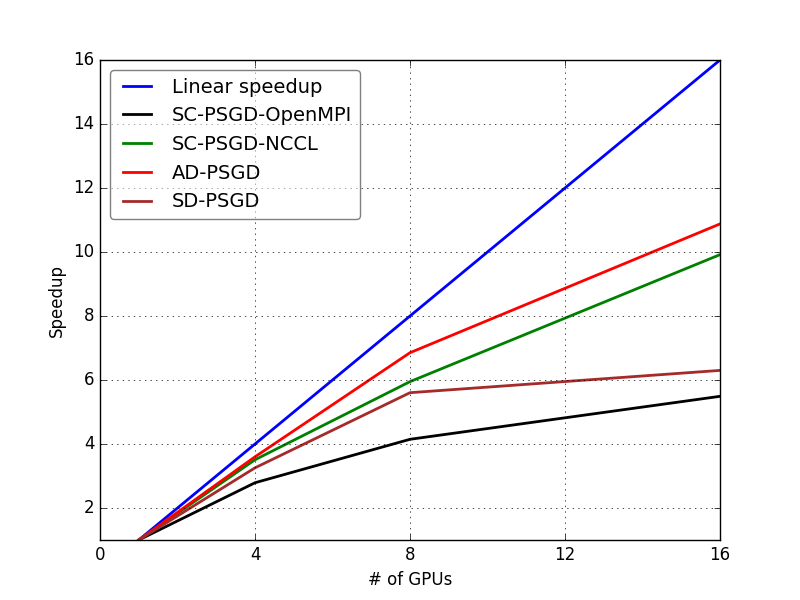}
      \caption{speedup}
  \end{subfigure}
  \caption{Heldout loss and speedup of investigated strategies.}\label{fig:loss_speedup}\vspace{-0.7cm}
\end{figure}

Fig.\ref{fig:loss_speedup} shows the heldout loss of the distributed strategies using 16 GPUs in the left panel. The total batch size is 2560 and therefore each learner has a local batch size of 160. The batch size 2560 is determined based on WERs. The initial learning rate is 0.1 which is the same as that of the baseline. In the first 10 epochs, it linearly warmed up to 1.0 after which it is annealed by $\frac{1}{\sqrt{2}}$ every epoch.  All three distributed training strategies converge to similar loss close to that of the baseline. Their WERs are also close to those of the baseline. Specifically, WERs are 7.6\% on SWB and 13.1\% on CH under \textbf{SC-PSGD}, 7.6\% on SWB and 13.3\% on CH under \textbf{SD-PSGD} and 7.6\% on SWB and 13.2\% on CH under \textbf{AD-PSGD}. On the other hand, their speedup performance differs significantly, which is presented in the right panel of Fig.\ref{fig:loss_speedup} as a function of the number of GPUs. The synchronous SGD (\textbf{SC-PSGD} and \textbf{SD-PSGD}) has a smaller speedup than the asynchronous SGD (\textbf{AD-PSGD}) due to the idle time of the learners in the synchronization. We also compare the impact of two implementations of allreduce on \textbf{SC-PSGD}: one is the open-source MPI allreduce (SC-PSGD-OpenMPI) and the other is the NCCL allreduce (SC-PSGD-NCCL). The latter is a faster allreduce implementation than the former, which improves the speedup for synchronous SGD. Since \textbf{SD-PSGD} uses partial model averaging, it is implemented with OpenMPI. The partial model averaging reduces the communication cost and therefore gives better speedup over SC-PSGD-OpenMPI. Among these strategies, the best speedup performance is given by \textbf{AD-PSGD} which achieves 11x speedup over 16 GPUs.

\begin{figure}[tbh]
   \centering
   \includegraphics[width=7cm, height=5cm]{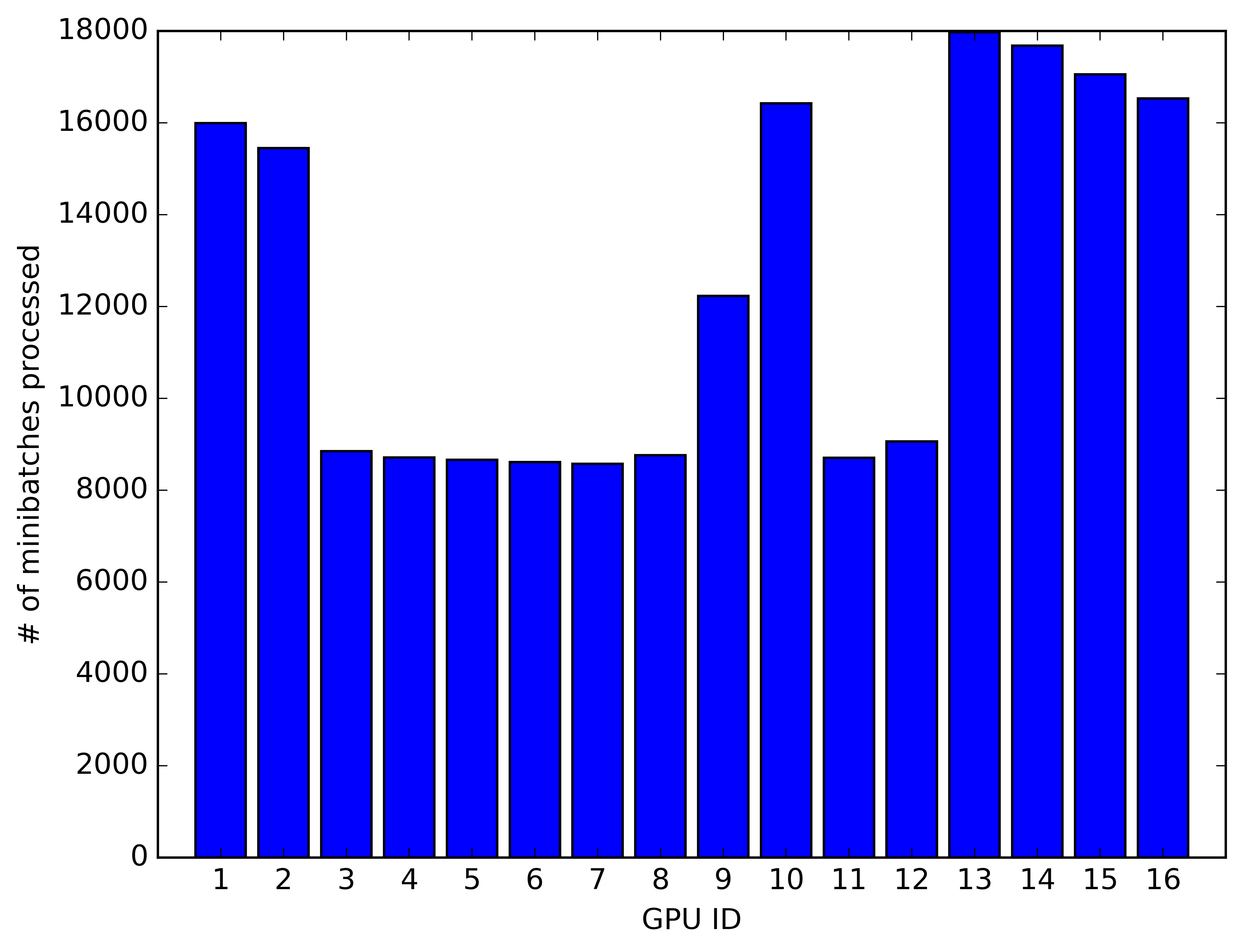}
   \caption{Distribution of workload on 16 GPUs for asynchronous decentralized parallel SGD.}\label{fig:load}\vspace{-0.7cm}
\end{figure}

One of the advantages of asynchronous SGD over synchronous SGD is its automatic load balancing. It allows a faster learner to consume more computation than a slow learner. To demonstrate this, we design a scenario in which 8 of the 16 GPUs in the cluster share running jobs from other tasks, which results in slow learners. Fig.\ref{fig:load} shows the distribution of the workload under $\textbf{AD-PSGD}$ across 16 GPUs in terms of the processed mini-batches in one epoch. It clearly shows the pattern that faster learners pick up higher workload during the training to create faster overall training. Furthermore asynchronous SGD can eliminate the ``straggler" problem that limits synchronous SGD. To show this effect, we design another scenario in Table \ref{tab:speedup_slow_learner} where we purposely slow down one GPU learner by 2x, 10x and 100x to make it a ``straggler". As can be seen from the table, this ``straggler" leads to significant prolonged training time in one epoch in synchronous SGD while not affecting asynchronous SGD. $\textbf{AD-PSGD}$ in this case delivers consistent speedup.

\begin{table}[tbh]
\centering
\begin{tabular}{l|c|c|c|c} \hline
\multirow{2}{*}{slow GPU learner} & \multicolumn{2}{c|}{\textbf{SC-PSGD}}  & \multicolumn{2}{c}{\textbf{AD-PSGD}}       \\ \cline{2-5}
                 &    hr/epoch   &    speedup    &  hr/epoch   &  speedup     \\ \hline\hline
no slowdown      &     1.09      &     8.70      &    0.87     &   10.88      \\ \hline
2x               &     1.67      &     5.71      &    0.89     &   10.63      \\ \hline
10x              &     6.24      &     1.52      &    0.91     &   10.42      \\ \hline
100x             &     57.73     &     0.16      &    0.92     &   10.38      \\ \hline
\end{tabular}
\caption{Runtime and speedup comparison on 16 GPUs when one GPU slows down by 2x-100x.}\label{tab:speedup_slow_learner} \vspace{-0.7cm}
\end{table}

In the second set of experiments, we try to show how the discussed training stategies can substantially help to shorten the training time on SWB2000 without sacrificing recognition accuracy, especially their scaling-out capability when increasing the number of GPUs. To maximize the parallelization performance it is critically important to increase the batch size while sustaining a good convergence. It was found that \textbf{AD-PSGD} can tolerate much larger batch size than its synchronous centralized counterpart \cite{Zhang_DDLASR}.

\textbf{HPC Setting} \ \  We use a 8-server cluster equipped with 1TB main memory and 8 V100 GPUs on each server. Each server has 2 9-core Intel Xeon E5-2697 2.3GHz CPUs.  Between servers are 100Gbit/s Ethernet connections. GPUs and CPUs are connected via PCIe Gen3 bus, which has a 16GB/s peak bandwidth in each direction. In order to maximize the feasible batch size and meanwhile effectively reduce the required communication bandwidth, we employ a hierarchical-ring (H-ring) configuration following \cite{Zhang_DDLASR}. In this configuration, GPU learners on the same computing node run \textbf{SC-PSGD} with a local ring by NCCL allreduce. They are called a super learner. All the super learners then form another ring running \textbf{AD-PSGD}. Therefore, it is a hierarchical implementation of synchronous and asynchronous SGD in one configuration.

Table \ref{tab:adpsgd_scaleout} shows the speedup and recognition performance.\footnote{In order to make the speedup comparable, we optimize the batch size against WER on 64 GPUs and then scale down to 32 and 16 GPUs with the same local batch size on each learner (128 per learner). Therefore, the batch sizes on 16 and 32 GPUs may not be optimal under these two conditions.} While training the LSTM acoustic model on SWB2000 with a single V100 GPU takes 195 hours,  it takes 20 hours on 16 V100 GPUs,  9.9 hours on 32 V100 GPUs and 5.2 hours on 64 V100 GPUs. This is equivalent to about 38x speedup with similar WERs. To the best of our knowledge, this is the best speedup reported on SWB2000 with this level of recognition accuracy by the time of submission of this paper.


\begin{table}[tbh]
\centering
\begin{tabular}{l|c|c|c|c|c} \hline
\multirow{2}{*}{GPUs} &  \multirow{2}{*}{batch size} & \multirow{2}{*}{training time} & \multirow{2}{*}{speedup} & \multicolumn{2}{c}{WER} \\ \cline{5-6}
                     &             &               &          &   SWB    &   CH    \\ \hline\hline
single V100 GPU      &    256      &    195 hr     &    -     &  7.5     &   13.0  \\ \hline
16 V100 GPUs         &   2048      &    20.0 hr    &  9.8    &  7.5     &   13.2  \\ \hline
32 V100 GPUs         &   4096      &    9.9 hr     &  19.7    &  7.5     &   13.2  \\ \hline
64 V100 GPUs         &   8192      &    5.2 hr     &  37.5    &  7.6     &   13.2  \\ \hline
\end{tabular}
\caption{Scaling-out performance on SWB2000 using the H-ring configuration with various numbers of GPUs.}\label{tab:adpsgd_scaleout}\vspace{-0.7cm}
\end{table}

\section{Summary}
\label{sec:sum}

In this article, we walked through the distributed training of DNN acoustic models using mini-batch based data parallel SGD. We gave an overview of existing distributed training strategies (synchronous vs. asynchronous, centralized vs. decentralized) in the speech community and analyzed their advantages and disadvantages.  We also studied their convergence and speedup performance on the popular public benchmark SWB2000 dataset. For distributed training using data parallel SGD, a batch size that is sufficiently large is a necessary condition for good speedup. It depends on optimization algorithms and a careful design of learning schedules. In addition, handling the interplay between communication and computation is crucial for high-performance distributed training. In practice, we strive for the maximum overlap between communication and computation when designing and implementing a distributed training strategy from both algorithmic and HPC architectural perspectives.

\bibliographystyle{IEEETran}
\bibliography{dist_asr}

\end{document}